\documentclass[twocolumn]{openjournal}
\usepackage{xcolor}
\usepackage{textgreek}
\usepackage[utf8]{inputenc}
\usepackage[english]{babel}

\usepackage{hyperref}
\hypersetup{
    unicode, 
    colorlinks=true,
    linkcolor=linkcolor,
    citecolor=linkcolor,
    filecolor=linkcolor,
    urlcolor=linkcolor,
}
\usepackage{color,colortbl}
\definecolor{linkcolor}{rgb}{0.0,0.3,0.5}
\usepackage{tensind}
\tensordelimiter{?}
\DeclareGraphicsExtensions{.bmp,.png,.jpg,.pdf}
\usepackage{verbatim}
\usepackage[normalem]{ulem}
\usepackage{orcidlink}
\usepackage{soul}

\graphicspath{ {./figs/} }

\begin{document}
\title{On the Use of WGANs for Super--Resolution in Dark-Matter Simulations\vspace{-1cm}}

\author{John Brennan$^{1,*}$\orcidlink{0000-0002-4428-6798}}
\author{Sreedhar Balu$^{2, 3, 4}$\orcidlink{0000-0002-5281-5151}}
\author{Yuxiang Qin$^{5}$\orcidlink{0000-0002-4314-1810}}
\author{John Regan$^1$\orcidlink{0000-0001-9072-6427}}
\author{Chris Power$^{6, 4}$\orcidlink{0000-0002-4003-0904}}
\thanks{$^*$E-mail:john.brennan@mu.ie}
\affiliation{
$^1$Centre for Astrophysics and Space Science Maynooth, Department of Physics, Maynooth University, Maynooth, Ireland \\ 
$^{2}$School of Physics, University of Melbourne, Parkville, VIC 3010, Australia\\
$^3$Facultad de Físicas, Multidisciplinary Unit for Energy Science, Universidad de Sevilla, 41012, Seville, Spain\\
$^4$ARC Centre of Excellence for All Sky Astrophysics in 3 Dimensions (ASTRO 3D)\\
$^5$Research School of Astronomy and Astrophysics, Australian National University, Canberra, ACT 2611, Australia\\
$^6$International Centre for Radio Astronomy Research (ICRAR), M468, University of Western Australia, 35 Stirling Hwy, Crawley, WA 6009, Australia
}

\begin{abstract}
    \noindent Super-resolution techniques have the potential to reduce the computational cost of cosmological and astrophysical simulations. This can be achieved by enabling traditional simulation methods to run at lower resolution and then efficiently computing high-resolution data corresponding to the simulated low-resolution data. In this work, we investigate the application of a Wasserstein Generative Adversarial Network (WGAN) model, previously proposed in the literature, to increase the particle resolution of dark-matter-only simulations. We reproduce prior results, showing the WGAN model successfully generates high-resolution data with summary statistics, including the power spectrum and halo mass function, that closely match those of true high-resolution simulations. However, we also identify a limitation of the WGAN model in the form of smeared features in generated high-resolution data, particularly in the shapes of dark-matter halos and filaments. This limitation points to a potential weakness of the proposed WGAN-based super-resolution method in capturing the detailed structure of halos, and underscores the need for further development in applying such models to cosmological data.
\end{abstract}


\maketitle

\section{Introduction}
    \noindent Super-resolution (SR) refers to the task of artificially increasing the resolution of data. The computer science community has worked on this problem for several decades and, as a result, developed several families of techniques for computing high-quality solutions to the problem \citep[e.g.,][]{Lepcha_2023}. Although the initial use case for most of these techniques was image enhancement, all of them can, in principle, be used to improve the resolution of physics simulations \citep[e.g.,][]{Fukami_2019, Xie_2017, Hong_2021}. Instead of enhancing the red, green and blue (RGB) channels of a two-dimensional image, one can think about enhancing various field channels (e.g., a density, temperature, velocity component etc.) of a three-dimensional simulation snapshot. Of particular interest to the authors are applications of these techniques to improving the resolution of dark-matter cosmological simulations.

    The main reason to consider SR techniques for astrophysics simulations is their potential to significantly reduce the computational cost of producing high-resolution (HR) data. While HR datasets are crucial for theoretical work in many areas of astrophysics, the simulations needed to produce them are computationally expensive. More broadly, fast approximation techniques for structure formation simulations are essential for observational cosmology surveys. These surveys require numerous realizations of cosmic large-scale structures on Gigaparsec scales, while also capturing the distribution of galaxies and their satellites on scales ranging from tens to hundreds of kiloparsecs \citep[e.g.,][]{Pujol_2017,Ding_2022}. Thus, a suitable balance between resolution and computational cost regularly needs to be found in order to produce meaningful results while avoiding impractical or intractable simulations -- a task which is not always feasible. SR methods offer a potential approach to bridge the gap between low-resolution (LR) and HR simulations, allowing researchers to run cheaper LR simulations and recover HR data from them. For instance, while not quite SR, similar methods described in \cite{Jacobus_2024} allowed the authors to make HR corrections to LR simulations without increasing its resolution. This enabled them to produce high-quality data representing a gigaparsec hydrodynamic volume, far beyond the capabilities of traditional simulation methods running on current supercomputers.
    
    In a trilogy of papers, \citep{Li_2021, Ni_2021, Zhang_2024} proposed a method based on a generative adversarial neural network (GAN) \citep{Goodfellow_2014} for enhancing the particle resolution of dark-matter-only simulations. Their approach involved converting particle position data into a displacement field with respect to an initial grid. A Wasserstein GAN model (WGAN) \citep{Arjovsky_2017}, inspired by a popular design known as styleGAN \citep{Karras_2019, Karras_2020} for image generation, was then used to enhance the resolution of this displacement field. The enhanced displacement field is then converted back to particle position data with a higher particle count. Their models achieved up to a 512-fold increase in particle resolution across a range of redshifts and reproduced several key statistical properties of high-resolution simulations, including power spectra, halo mass functions, and halo merger trees.
    
    In this work, we reproduce earlier results for modest super-resolution enhancement factors and highlight a key limitation of the WGAN-based approach in predicting high-resolution data from corresponding low-resolution inputs. Specifically, we find that the model struggles to accurately reproduce the shapes of dark-matter halos and filaments, leading to noticeable discrepancies when compared with true high-resolution simulations. We further investigate this issue and present initial results from efforts to mitigate the problem.
    
    The paper is structured as follows: in section \ref{sec:method} we introduce the methods along with the datasets used, and section \ref{sec:results}  lists our main results. We summarise and conclude in section \ref{sec:conclusion}. Throughout we use a WMAP9 cosmology \citep{Hinshaw_2013}: $(\Omega_{m}, \Omega_{\Lambda}, \Omega_b, h) = (0.276, 0.724, 0, 0.703)$.

%--------------------------------------------------------------------
\section{Method}\label{sec:method}
    \noindent In this section we give details on the neural networks used in this work, and both the datasets and procedure used to train them. We closely follow the methods outlined in \cite{Li_2021} and additionally use a generalization of one of the models, as discussed below. For the purpose of testing the SR approach, we focus on enhancing the spatial resolution of data by a factor of 2 (that is, the spatial resolution of displacement fields is increased by a factor of 2 whereas the number of particles is increased by a factor of 8) rather than a factor of 8 as used by \cite{Li_2021}. Focusing on a smaller scale factor enables us to train neural network models using fewer computational resources while still allowing us to analyze the accuracy of these models in recreating HR data.

    \subsection{Dataset Details}\label{sec:dataset}
    \noindent The training, validation and test datasets used in this work were created from a set of dark-matter-only simulation pairs, all of comoving volume $(100~{h}^{-1}\mbox{Mpc})^3$. Each simulation pair consists of an LR and an HR simulation of $64^3$ and $128^3$ particles respectively, with corresponding particle masses of $4.2 \times 10^{11} M_\odot$ and $5.2 \times 10^{10} M_\odot$. Initial conditions were generated using the MUSIC2--monofonIC\footnote{bitbucket.org/ohahn/monofonic/src/master/} initial conditions generator \citep{MONOFONIC:3LPT, MONOFONIC:BARYONS}. For each simulation pair, we fixed the random seed used to generate their initial conditions. Each pair therefore represents both a low- and a high-resolution realization of the same cosmological volume. All our simulations were run from redshift $z = 99$ down to $z=0$ with a gravitational softening length of 1/30th the mean inter-particle separation using the SWIFT simulation code \citep{Schaller_2024}. Snapshots were taken at redshift 0 and a set of 16 simulation pairs was used to create the training dataset while a single simulation pair each was used for both validation and test datasets.

    All datasets consisted of pairs of LR and HR patches of dark-matter displacement fields. These were created by first converting the position data, of dark-matter particles in each snapshot, into displacement fields. This was achieved by computing each particle's displacement from its initial grid position. Displacement fields were stored as multi-dimensional arrays, or tensors, with a shape $(3, N, N, N)$ where the first dimension corresponds to the $x$, $y$ and $z$ components, or channels, of the displacement field. The next three dimensions correspond to the $x$, $y$ and $z$ indices of points on the $N\times N\times N$ regular grid. Here, $N$ is 64 and 128 for the LR and HR patches respectively. These arrays were then segmented into patches, yielding a set of 64 LR patches of shape $3\times20^3$, and $3\times32^3$ HR counterparts, for each snapshot. This resulted in $64\times 16=1024$ LR-HR patch pairs in the training set, and 64 patch pairs in both the validation and test sets. We take the LR patches to be $3\times 16^3$ patches with 2 cells of padding on each side, making the total patch size $3\times20^3$. Correspondingly, the $3\times 32^3$ HR patches should be taken as HR copies of the inner $3\times 16^3$ regions of the LR patches. Both the HR patches and the inner $16^3$ regions of the LR patches represent cubes with a spatial length of $25\ \rm{h^{-1}}\mbox{Mpc}$ (comoving). Random permutations of the $\rm{xyz}$-axes were also used as a data augmentation technique to artificially expand the dataset size during training.
    
    \subsection{Neural Network Details}
    \noindent Following the approach of \cite{Li_2021}, a Wasserstein generative adversarial neural network model with a gradient penalty (WGAN) \citep{Arjovsky_2017, Gulrajani_2017} was used to enhance dark-matter data. WGANs are widely regarded as a more stable and robust alternative to classical GAN models, mitigating common issues such as mode collapse and training instability. They consist of two neural network models: a generator model and a critic, or discriminator, model. The generator model is responsible for producing synthetic HR versions of any LR data passed to it, while the critic model is used to assign a score to any HR data it is given. The score assigned by the critic model should be taken as a measure of how much the data resembles real HR simulation data. Below we outline the neural network architectures we used for both of these models.

    \subsubsection{Generator Model}
\begin{figure*}
        \centering
        \includegraphics[width=0.9\linewidth]{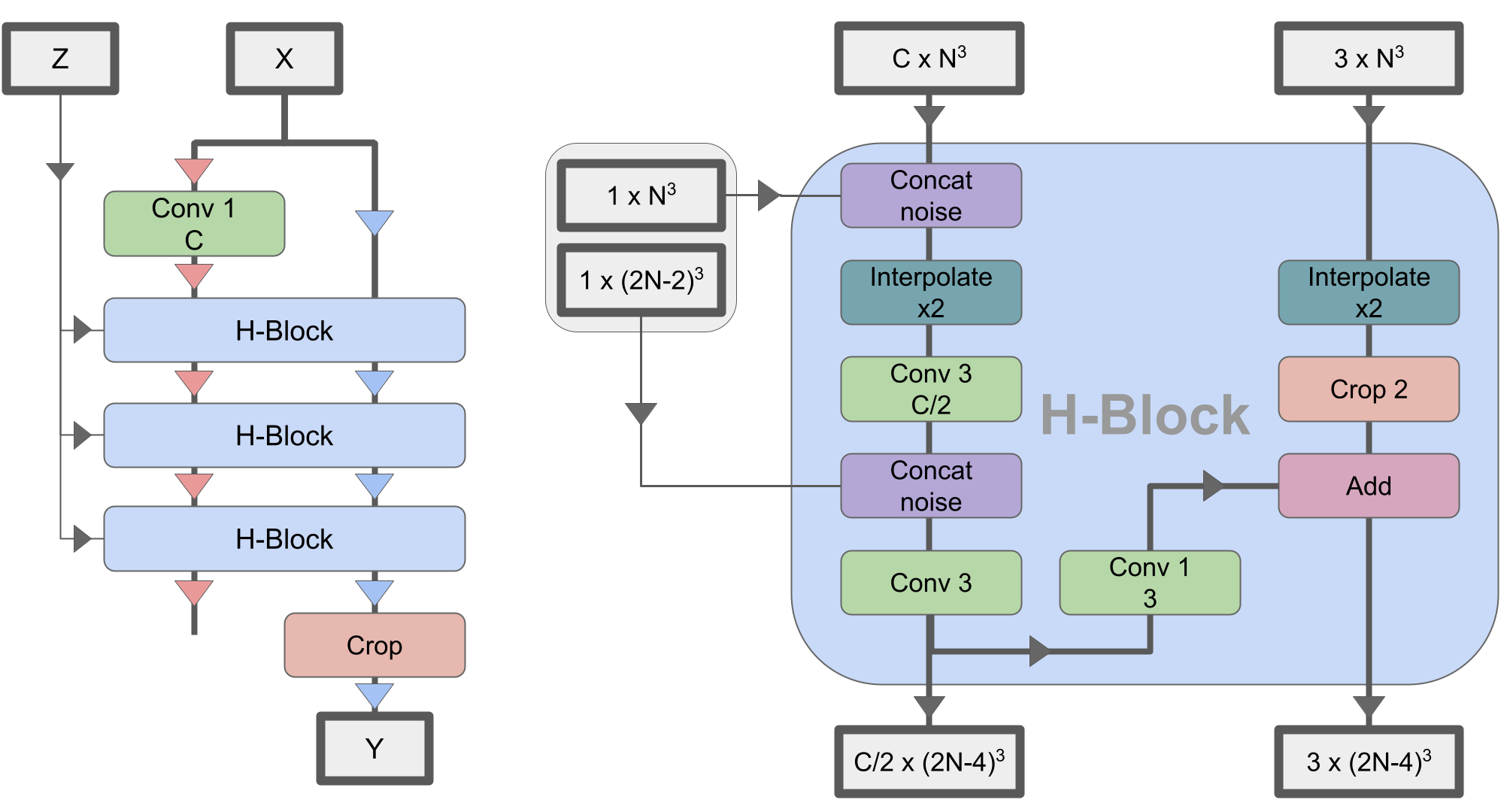}
        \caption{\normalsize The graphics above illustrate the tensor operations used by the genertaor model. Tensors are denoted by gray boxes with thick borders, with their shape indicated inside. All tensors are rank 4 and have a shape $C\times N^3$ where $C$ is the number of data channels and $N$ is the number of cells in each of the $\rm{xyz}$ directions. Operations on tensors are shown as rounded boxes with the name of the operation inside. Key parameters are also shown in the these boxes: the kernel size for convolutional layers, the crop size for crop operations and the scale factor for trilinear interpolation operations. If a convolutional layer changes the number of channels, the new number of channels is also given. The generator architecture is depicted on the left. The input tensor $X$ (LR data), is passed to both a primary and an auxiliary data stream, indicated by blue and red arrows respectively. Data in both streams are processed and upscaled by a sequence of H-blocks. The auxiliary stream is also processed by an initial convolutional block to expand the number of channels in the stream. The output of the primary stream is cropped and taken as the output $Y$ of the model. Each H-block additionally takes two tensors, containing Gaussian noise, as input. The collection of all such tensors is denoted by $Z$ and constitutes the latent variable for the model. The graphic on the right shows the internal structure of a H-block.}
        \label{fig:generator}
    \end{figure*}

    \noindent The generator model consists of an initial convolutional block and a sequence of blocks, known as H-blocks (described below), to upscale LR data and synthesize new HR details. Each H-block has two main inputs and two corresponding outputs. A sequence of H-blocks composed together thus forms a ladder shaped architecture with two parallel data streams forming the legs of the ladder as depicted in Fig. \ref{fig:generator}. We refer to these data streams as the primary and auxiliary data streams. The primary data stream processes data tensors with three channels, corresponding to the $x$, $y$ and $z$ components of the displacement field. The auxiliary data stream, on the other hand, processes data with a variable number of channels depending on the position of the H-block in the model. These channels correspond to different feature maps learned by the model for upscaling data. The auxiliary data stream starts with a fixed base number of channels which is halved after each H-block. The LR data to be upscaled is passed, as a $(3, N, N, N)$ tensor, directly to the primary data stream of the model while an initial convolutional block is used to increase the number of channels to the base number and create a $(C_{\mbox{base}}, N, N, N)$ input tensor for the auxiliary stream. We choose to use $C_{\mbox{base}}=64$ as the base number of channels for the auxiliary stream of our generator model. The output of the primary data stream is cropped to remove any excess spatial cells that may have been created by the H-blocks. The cropped output is then taken as the output of the generator model and should be taken as a HR counterpart of the LR data given to the generator.

    In addition to the two main inputs, each H-block also takes standard Gaussian noise as extra input. More precisely, when a H-block receives a primary input tensor with a shape $(3, N, N, N)$, two additional tensors are provided as input: one with shape $(1, N, N, N)$ and another with shape $(1, 2N-2, 2N-2, 2N-2)$. Both of these extra tensors contain samples drawn from a standard normal distribution. This noise is used by the H-block to synthesize new HR details to be added to the upscaled LR data. In keeping with the terminology used in the literature on GANs, the collection of all such noise tensors passed to the H-blocks is referred to as a sample $z$ from a latent space $Z$ for the generator. Denoting the generator by $G$ and the LR data to be upscaled by $X$, then for a given sample $z$ from the latent space, the enhanced data can be written as $Y = G(X, z)$. Furthermore, for two different samples from the latent space $z_1$ and $z_2$, the generator outputs $Y_1 = G(X, z_1)$ and $Y_2 = G(X, z_2)$ can be taken as non-identical HR realizations of the same LR input $X$. These HR realizations differ in the small-scale details, particularly in regions where the LR data underdetermines the HR structure, while remaining consistent with the large-scale structure set by LR input.

    \begin{figure*}
        \centering
        \includegraphics[width=0.9\linewidth]{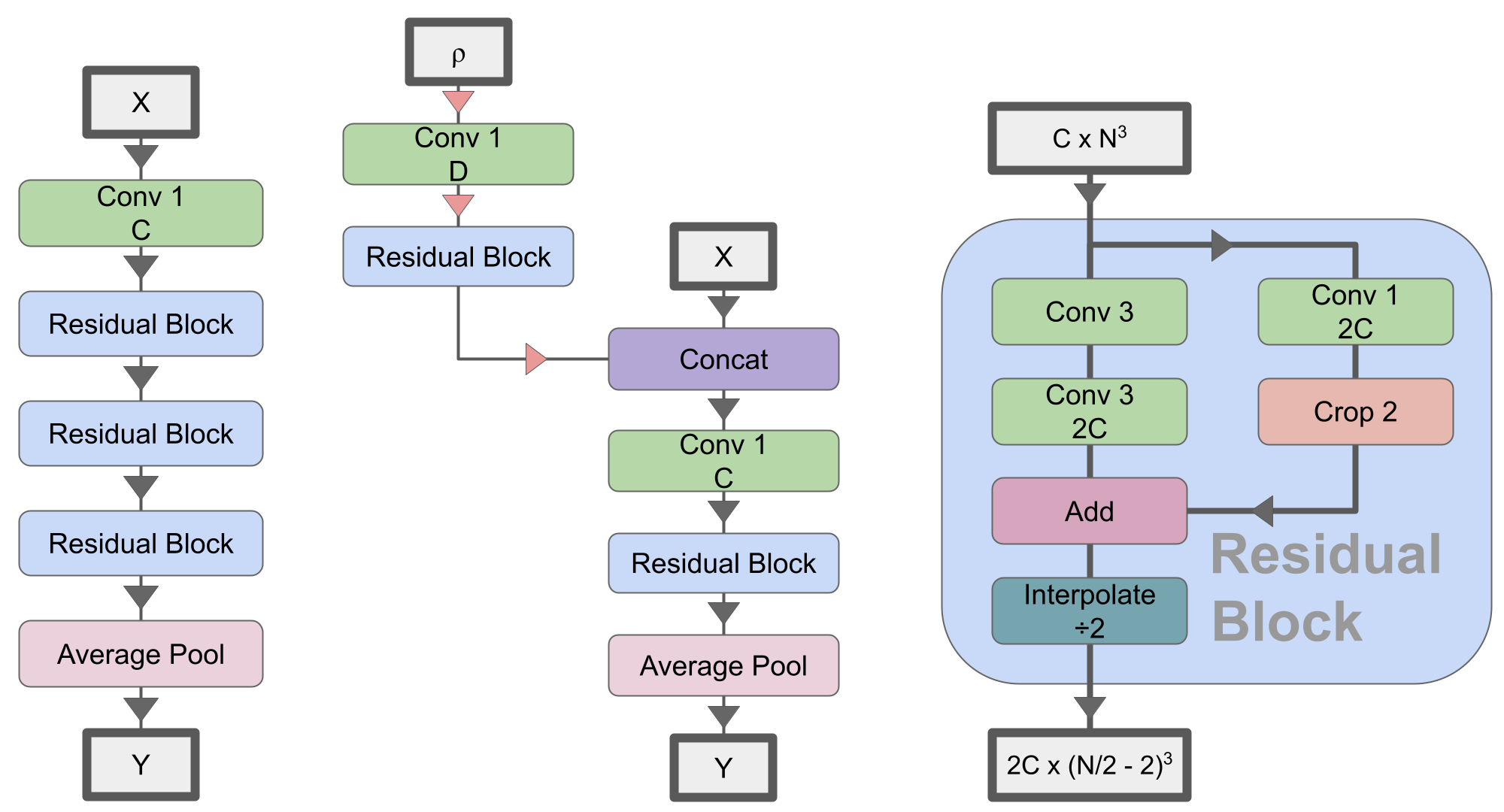}
        \caption{\normalsize The above graphics use the same graphical notation used in Fig. \ref{fig:generator}. The architecture of the critic neural network model is shown on the left while the architecture of a generalized critic model is shown in the center. The generalized model processes density field data in its own branch (indicated by red arrows) before being merged into the main branch (indicated by dark gray arrows). The number of residual blocks in each branch is determined by the size of the input data. The architecture of the residual block is shown on the right.}
        \label{fig:critic}
    \end{figure*}

    To outline the architecture of the H-block layer (depicted in Fig. \ref{fig:generator}), we note it is responsible for two key tasks. Firstly, it increases the resolution of data in both the primary and auxiliary streams. Data tensors with spatial size $N$ are upscaled, by a single H-block, to a size $2N-4$. This is done in the auxiliary stream using a trilinear interpolation operation (linear interpolation in three dimensions) to increase the spatial dimensions by a factor of 2, followed by a pair of convolutional layers which remove 1 cell of padding each from the edges of the tensor. In the primary data stream, a trilinear interpolation operation is again used to increase the spatial dimensions by a factor of 2 and then a crop operation removes 2 cells of padding from the edges of the tensor to keep the spatial dimensions in both streams aligned. Secondly, the H-block injects noise into the feature channels of the auxiliary stream. This is done by concatenating noise with the data in the auxiliary stream at two points: before the interpolation operation and before the second convolutional layer as depicted in Fig. \ref{fig:generator}. The noise added to the data is the same noise passed to the H-block as input as discussed above. Finally, a convolutional layer performs a projection operation that maps the auxiliary data into a 3-channel tensor. This tensor is then added to the primary data before being returned as output. We can think of this operation as adding any HR details synthesized by the auxiliary stream to the primary output.

    Each H-block both increases the spatial dimensions of the input data given to its primary and auxiliary streams and halves the number of channels used by the auxiliary stream. More precisely, for auxiliary and primary inputs with shapes $(C, N, N, N)$ and $(3, N, N, N)$ respectively, the auxiliary and primary outputs of the H-block have shapes $(C/2, 2N-4, 2N-4, 2N-4)$ and $(3, 2N-4, 2N-4, 2N-4)$ respectively. The number of H-blocks used in the generator model thus determines the factor by which LR data is enhanced. As mentioned, in this work we focus on enhancing LR data by a factor of 2. We therefore use a single H-block in our generator model and pad our LR data with 2 cells on each side to account for the boundary cells consumed by the convolutional layers in the H-block. The spatial dimensions of the generator output were cropped on each side by 2 cells.

    \begin{figure*}
        \centering
        \includegraphics[width=1\linewidth]{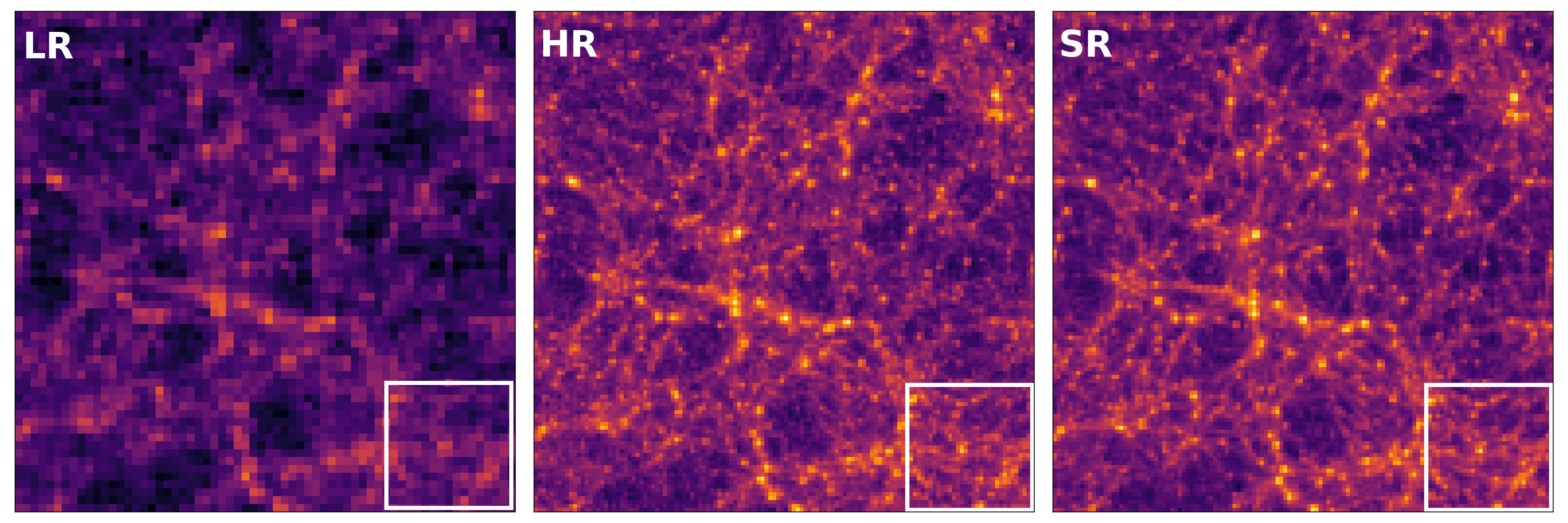}
        \includegraphics[width=1\linewidth]{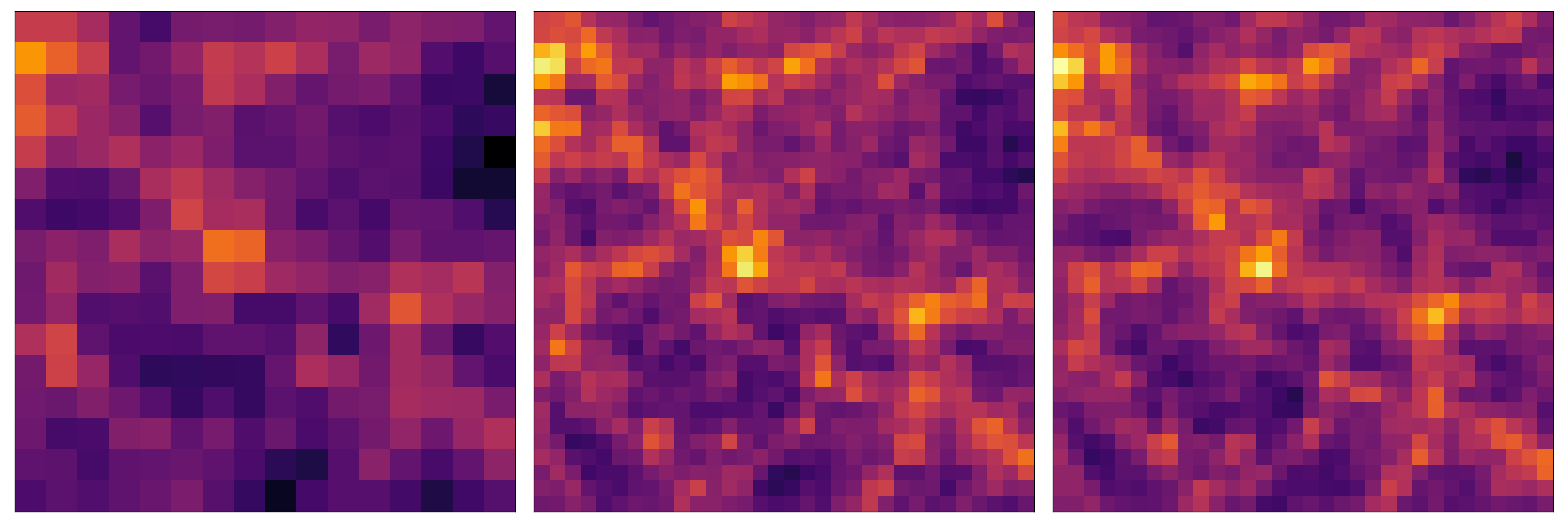}
        \caption{\normalsize Shown above are projected dark-matter density fields calculated using a cloud-in-cells method. The top row of images, from left to right, show density fields for a $z=0$ snapshot from low and high resolution simulations of a 100 $\rm{h^{-1}}$ Mpc box, together with an enhanced snapshot respectively. The enhanced snapshot was created using the low resolution snapshot as input to the model. The bottom row of images show zoom-ins of the 25 $\rm{h^{-1}}$ Mpc regions highlighted in the top row by enclosing boxes. \textit{Takeaway}: We see, even on scales where the resolution of the density field is apparent, it is difficult to distinguish between a high resolution density field and one created using the SR technique.}
        \label{fig:dark-matter-density}
    \end{figure*}
    
    \subsubsection{Critic Model}\label{sec:critic-model}
    \noindent To compute a score that measures how realistic a displacement field is, the critic model takes a number of inputs. Namely, in addition to the HR displacement field to be scored, the critic also takes a linearly upscaled copy of the corresponding region of the LR displacement field. Additionally, density fields are created from both of these displacement fields, using a cloud-in-cells method, and provided as input to the critic model. The critic then outputs a single number as a score for how realistic the given data is. Note, the density fields created from displacement fields are chosen to have the same spatial dimensions as those displacement fields. This means that the number of cells in the density field is equal to the number of particles in the HR data. This choice allows both the displacement fields and the density fields to be placed in the same input tensor with shape $(8, N, N, N)$. The number of channels here comes from the three $x$, $y$ and $z$ channels of both displacement fields and a single channel for each density field.
    
    The critic model first uses an initial convolutional layer to increase the number of channels of the input to a base number which we took as 64 for this work. Following the original design, a sequence of residual blocks is then used to downscale the spatial dimensions of data and simultaneously increase the number of its channels as shown in Fig. \ref{fig:critic}. Inspired by the discriminator architecture used by the patchGAN model \citep{Isola_2018}, the output of the last residual block has a spatial size greater than 1 and is interpreted as a tensor of scores for overlapping patches in the input. In particular, each element of this tensor is a score for its receptive field in the input of the critic. A global average pool operation is used to combine these patch scores into a single score for the critic's input.

    The main components of a residual block are three convolutional layers and a trilinear interpolation operation. As depicted in Fig. \ref{fig:critic}, the input to a residual block is passed to a pair of composed convolutional layers, with $3\times 3\times 3$ kernels, and to a separate convolutional layer with a $1\times 1\times 1$ kernel. The output of the single convolutional layer is then cropped before being added to the output of the first pair of convolutional layers. The data is then downscaled by a factor of two using a trilinear interpolation operation before being returned as output. The number of channels in the input data is also doubled by the convolutional layers. Hence, the residual block maps an input tensor of shape $(C, N, N, N)$ to an output tensor with shape $(2C, N/2-2, N/2-2, N/2-2)$. The critic model used in this work used two residual blocks to map input data with shape $(8, 32, 32, 32)$ to a tensor of scores of shape $(256, 5, 5, 5)$ before the global average pooling operation reduced this tensor to a single scalar.

    One generalization of the original critic model entails separating the density and displacement fields into two distinct input tensors. Doing so opens the possibility of using density fields with spatial dimensions larger than those of the corresponding displacement fields. The density field plays a crucial role in aiding the critic model as it contains information regarding the proximity of particles to neighboring particles and the power spectrum. It also provides important gradients to the generator model during training. Density fields with higher spatial resolution can therefore provide more valuable information to both models during training. By separating the density fields into their own input tensor for the critic, we can add additional layers to the critic to process the density fields separately before being merged with the displacement fields. Specifically, we can add an additional branch of residual blocks to the model that can process larger density fields before combining them with the displacement fields as depicted in Fig. \ref{fig:critic}. Each branch in such a model has a convolutional layer that changes the number of channels to a base number, which in Fig. \ref{fig:critic} we denote by $D$ for the density branch and $C$ for the main branch. We test such a model in addition to the original critic model and report results in the next section.
    
    \subsection{Training Details}
    \noindent Both the generator $G$ and critic $C$ were trained using a Wasserstein GAN training algorithm \citep{Arjovsky_2017} which we now outline. During training, the model is fed training data in randomized batches of size 4 until all the training data is used, at which point a new training epoch begins and the process is repeated. Recall that each batch consists of several (in this case 4) LR and HR pairs of displacement fields. For each batch of training data, both models are updated in separate steps which we call the critic step and the generator step. In these update steps, ADAM optimizers are used to incrementally update the parameters of each of the models. We refer to these optimizers as the critic and generator optimizers. The learning rates used for these optimizers were $1\times 10^{-5}$ and $2\times 10^{-5}$ respectively while $\beta_1$ was set to 0 and $\beta_2$ was set to $0.99$ for both.
    
    In the critic step, the generator $G$ is first used to create fake HR counterparts to the LR data $x$ in the training batch. To this end, latent variables $z$ are sampled from the generator's latent space and the fake HR data is taken as $X_G = G(x, z)$. Then, the LR data is upscaled to the same resolution as the HR data using trilinear interpolation. Density fields are then produced for all displacement fields (fake, real and interpolated) using a cloud-in-cells method. The data is then organized into two inputs for the critic model: one containing real HR data $X_{\mbox{real}}$ and the other containing fake HR data $X_{\mbox{fake}}$. As described in \ref{sec:critic-model}, these inputs also contain the interpolated LR data. The critic model is then used to compute scores for both real and fake data and the following loss value is computed:
    \begin{equation}\label{eqn:loss_function}
        L_{\mbox{critic}} = C(X_{\mbox{fake}}) - C(X_{\mbox{real}}).
    \end{equation}
    The above quantity is averaged over all data in the batch. A gradient of the averaged loss value, with respect to the parameters of the critic model, is then computed and used by the critic optimizer to update the critic parameters to reduce the above loss value. Additionally, every sixteen iterations of the critic step, a regularization term known as gradient penalty is also computed and added to the loss value before its gradient is calculated. The gradient penalty term serves to ensure that the critic model is Lipschitz continuous with respect to its input \citep{Gulrajani_2017}. Intuitively, after the critic step, the critic model should improve at assigning low scores to fake data and high scores to real data and thus be better at distinguishing between real simulation data and data produced by the generator model.

    Once the critic step completes, the generator step is performed. During the generator step, more latent variables $z$ are sampled from the latent space and the generator is used again to create fake HR counterparts to the LR data $x$ in the batch. This data is then prepared as input for the critic model, $X_{G(x, z)}$, by similarly pairing it with upscaled LR data and associated density fields as in the critic step. The following loss value is then computed, and its gradient is used by the generator optimizer to update the generator parameters:
    \[ L_{\mbox{generator}} = -C(X_{G(x)}). \]
    After this update step, the generator is expected to improve in generating data that receives a higher score from the critic model, and thus better resembles real HR data, as judged by the critic.

\section{Results}\label{sec:results}
    \noindent We now present results from enhancing the resolution of a snapshot from a dark-matter-only simulation. We first show that the model can produce data with expected summary statistics and then present a limitation of the approach: its failure to reproduce the morphology of dark-matter halos and filaments. Note that, while all models used in this work were trained on the training set discussed in \ref{sec:dataset}, the snapshot that was enhanced here was taken from the test set. That is, the simulation data used to produce the plots that follow were not used to train any of the models. We also note that all models used in this work were trained on a NVIDIA H100 GPU with 64GB of memory.

    \subsection{Enhanced Dark-Matter Snapshot}
    \noindent To increase the particle resolution of the LR snapshot, a displacement field was created from the particle position data and then segmented into displacement fields for cubic regions covering the entire cosmological box. This was done in the same way that the training set was created (see \ref{sec:dataset}). A trained generator model was then used to increase the resolution of each cubic region before assembling all enhanced regions into a full cosmological box. Our initial results were produced using a WGAN with the original critic model architecture described in section \ref{sec:critic-model}.
    
    In Fig. \ref{fig:dark-matter-density} we show the cloud-in-cells density field of the dark-matter distribution from an enhanced snapshot that was generated by a trained model as discussed above. We plot this beside similar plots for the LR snapshot used to generate it and a corresponding HR snapshot from a real simulation. We find the WGAN model is capable of generating particle data that produce high quality density fields that are hard to distinguish visually from that of real HR data.

    \begin{figure}
        \centering
        \includegraphics[width=1\linewidth]{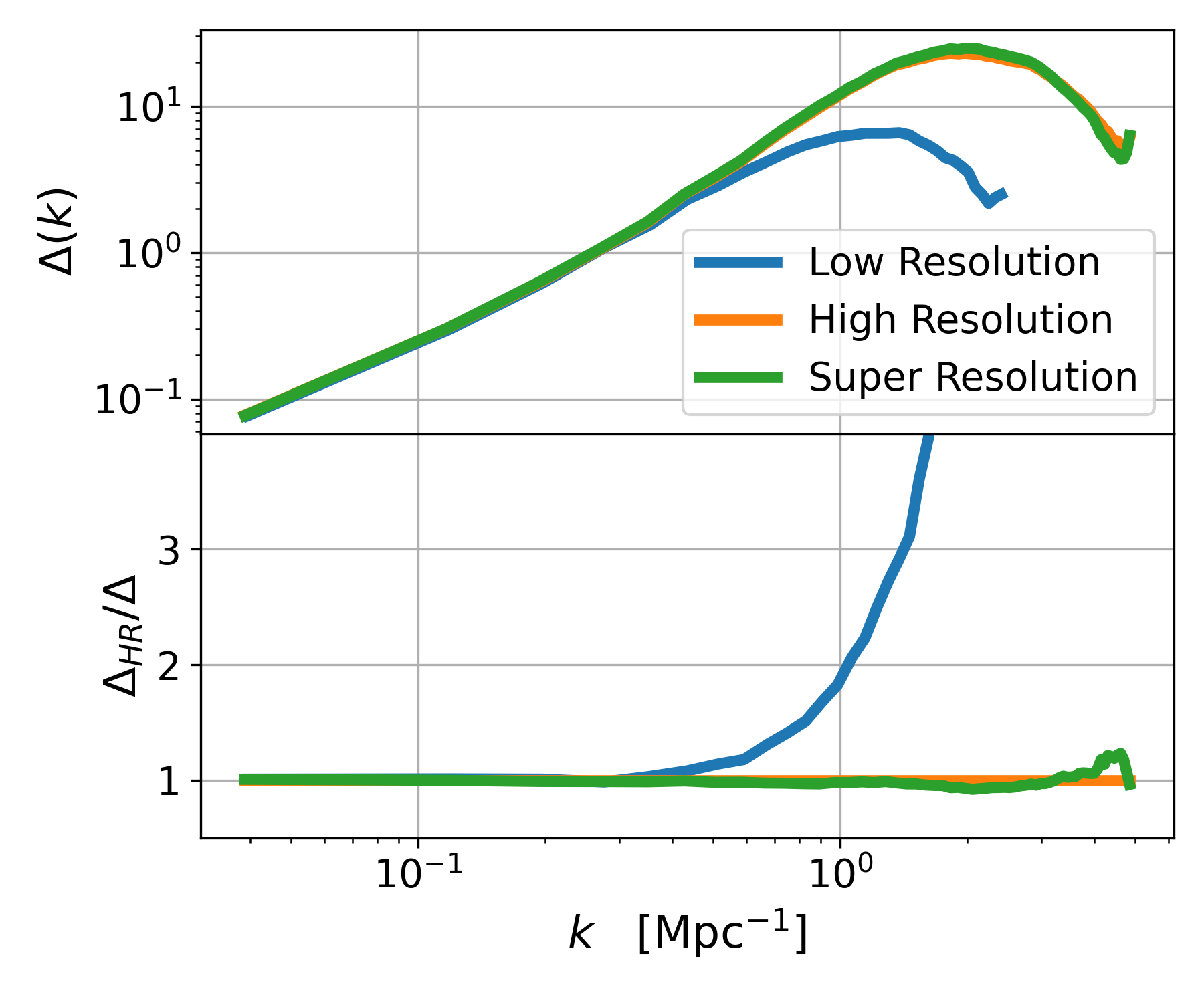}
        \caption{\normalsize The top panel shows the dimensionless power spectrum for the three snapshots shown in Fig. \ref{fig:dark-matter-density}. The bottom panel shows the ratio with respect to the HR power spectrum. \textit{Takeaway}: The power spectrum generated by the SR method shows a substantial improvement over the LR case and closely aligns with the HR power spectrum.}
        \label{fig:power-spectrum}
    \end{figure}

    \begin{figure}
        \centering
        \includegraphics[width=1\linewidth]{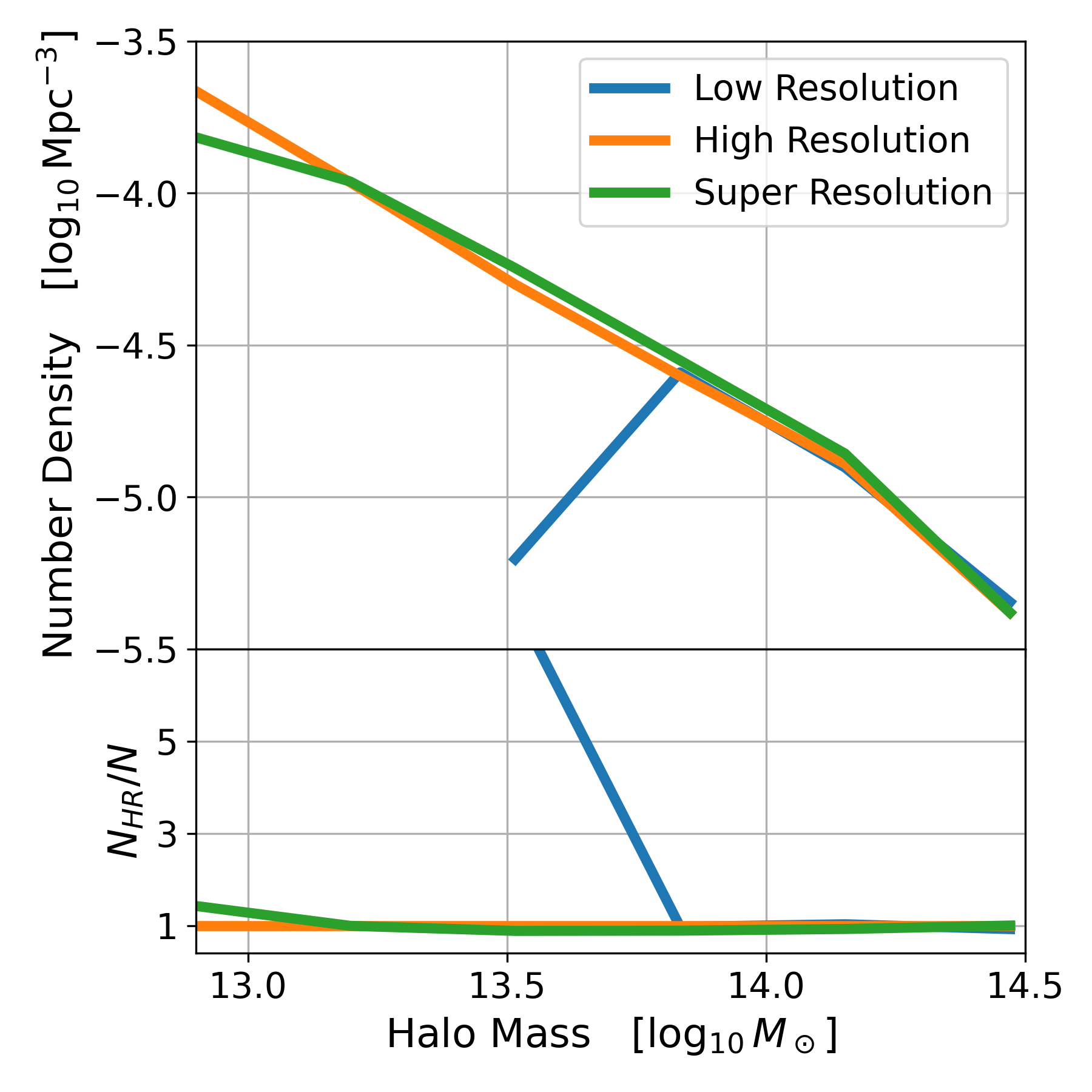}
        \caption{\normalsize The top panel shows the halo mass function for LR, HR and SR snapshots. The halo mass function was computed using a friends-of-friends algorithm, with a linking length of $0.2$ times the mean inter-particle spacing, to identify halos. The bottom panel shows the ratio with respect to the HR halo mass function. \textit{Takeaway}: The halo mass function of the enhanced snapshot closely matches that of the HR simulation, including at mass scales not resolved by the LR snapshot.}
        \label{fig:halo-mass}
    \end{figure}

    \begin{figure*}
        \centering
        \includegraphics[width=1\linewidth]{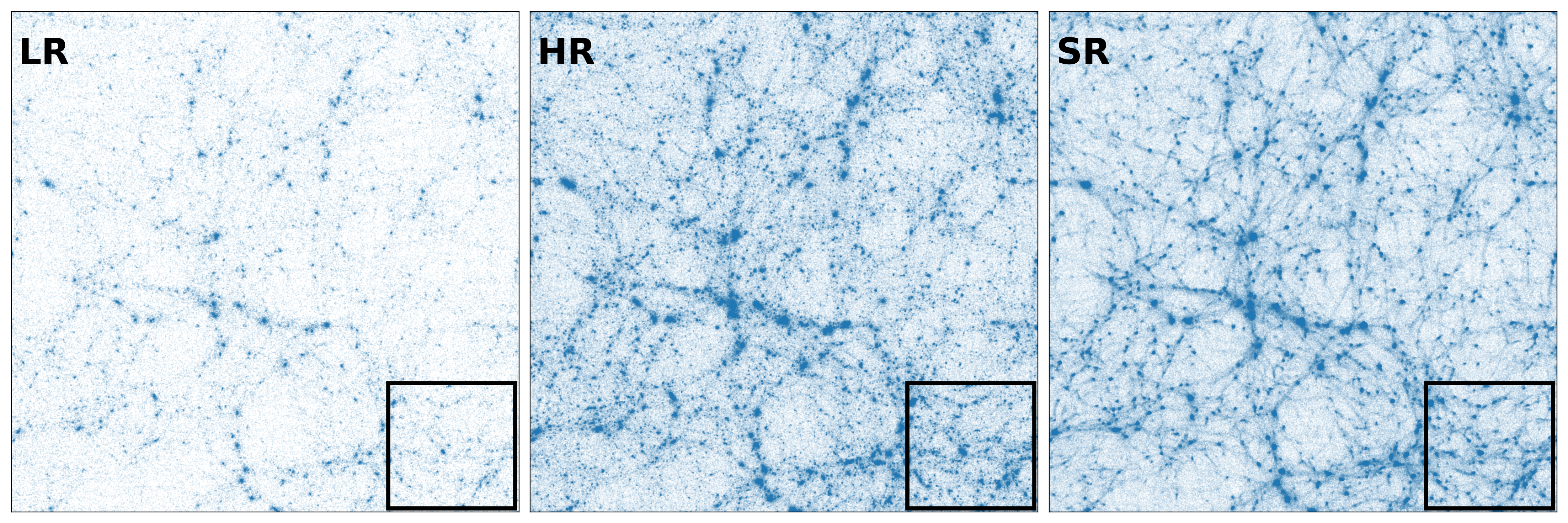}
        \includegraphics[width=1\linewidth]{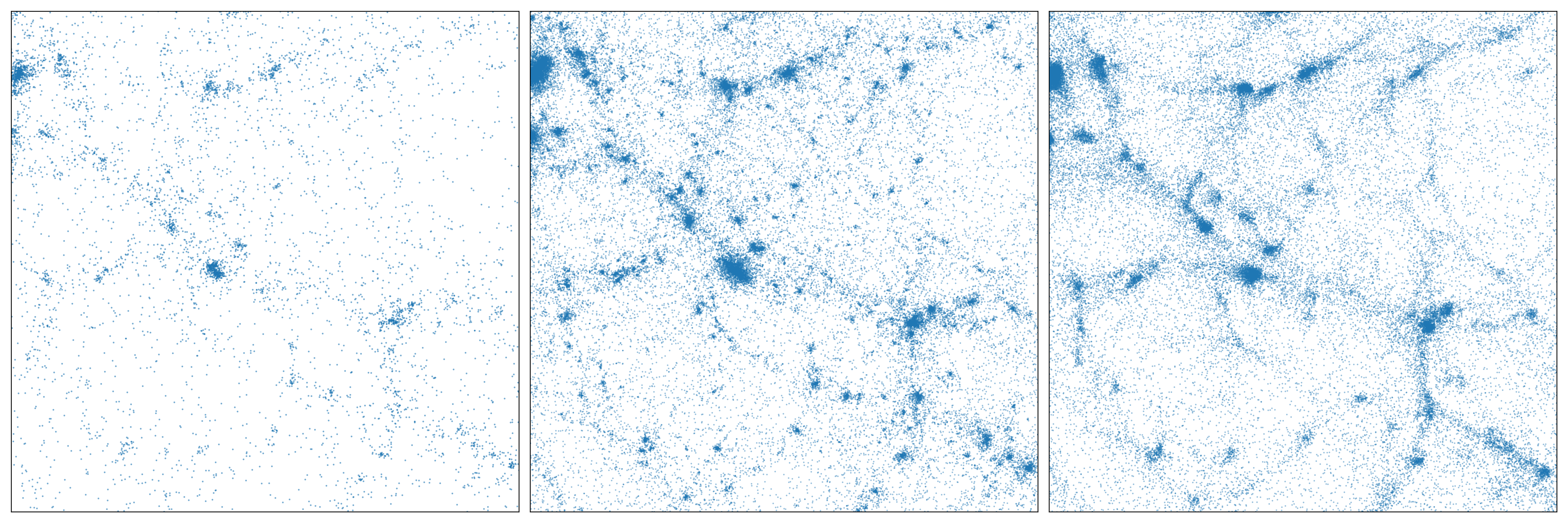}
        \caption{\normalsize Here we show a replica of Fig. \ref{fig:dark-matter-density} where particle scatter plots are used to show the dark-matter distributions instead of a projected density field. \textit{Takeaway}: Particularly in the zoom-in regions, we can clearly distinguish between HR and SR data by noting that the blotchy characteristics of the HR data are replaced in the enhanced data by smeared features.}
        \label{fig:dark-matter-scatter}
    \end{figure*}
    
    The enhanced snapshot also has summary statistics that align with that of a real HR snapshot. For instance, the power spectrum $\Delta(k)$ for all three snapshots are shown in Fig. \ref{fig:power-spectrum}. There, we see that the power spectra of all three snapshots agree at low-$k$ value (ie. large scales). More importantly, at high-$k$ (ie. small scales) the power spectrum of the enhanced snapshot diverges from that of the LR data and continues to agree with the HR data.
    
    The halo mass function for each snapshot was also computed using a friends-of-friends method \citep[e.g.,][]{Davis_1985} with a linking length of $0.2$ times the mean inter-particle spacing and by only considering halos with at least 100 particles. We similarly find that the enhanced data agrees with the HR data. The mass functions are shown in Fig. \ref{fig:halo-mass} and demonstrate that the enhanced snapshot successfully reproduces the abundance of halos across the same mass range as the HR data. In particular, the enhanced snapshot captures the higher abundance of low-mass halos, which are poorly resolved in the LR data, while maintaining accuracy for high-mass halos. These findings suggest that the SR displacement fields retain important statistical features of the target HR simulation.

    Enhancing LR snapshots with a trained generator model takes about 2 seconds on a single CPU \footnote{The CPU used was Intel Core i9-10940X CPU}. This is negligible compared to the wall-clock time for the LR simulation to reach redshift $0$ using the same CPU (approximately 7 minutes). While the wall-clock time for the HR counter part to reach the same redshift was 1 hour, generating HR snapshots with SR is significantly faster, with greater speed-ups expected when larger enhancement factors are used.
    
    \subsection{Morphology of cosmic structure}
    \noindent One major discrepancy we found between real and generated HR data is in relation to the shape of dark-matter halos. We found that halos in the data created by a trained generator tend to have smeared features compared to halos formed in real HR simulations. We find this discrepancy is most apparent in scatter plots of the particle positions. In Fig. \ref{fig:dark-matter-scatter} we show scatter plots of the same snapshots used to create the density fields in Fig. \ref{fig:dark-matter-density}. The smearing of halo shapes appears to occur on length scales below the spatial resolution of the density fields used during training. As a result, the smeared features are not resolved by the cloud-in-cells density fields used during training, meaning the neural network models may learn to ignore it. This discrepancy may limit the applicability of SR methods in analyses that are sensitive to small-scale halo morphology.

\begin{figure}
        \centering
        \includegraphics[width=1\linewidth]{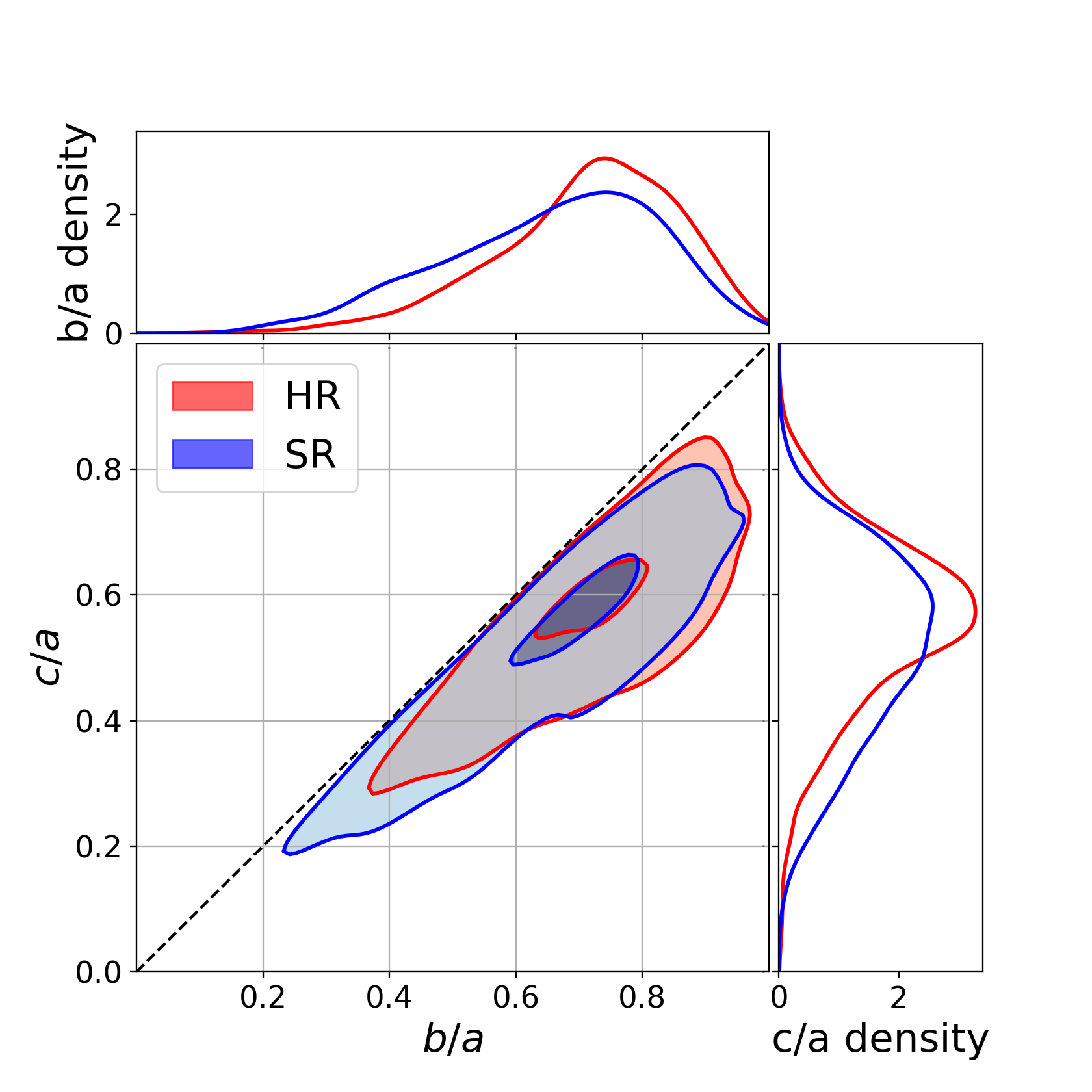}
        \caption{\normalsize Here we compare the distribution of shape parameters $(b/a, c/a)$ for both HR and SR halos. Contour plots of the two joint distributions are shown in the main panel while the marginal distributions are shown in the top and right hand panels. We find the distributions for SR halos are shifted towards lower values of $b/a$ and $c/a$ in comparison to the HR distributions. \textit{Takeaway}: SR halos are more elongated (or prolate) compared to HR halos.}
        \label{fig:shape-distribution}
    \end{figure}

    To quantify this discrepancy, we computed the shape of each halo using the method denoted $E1$ in \citep{Zemp_2011}. This method fits an ellipsoid to a given dark matter halo. We denote the lengths of the principle semi-axes of an ellipsoid by $a > b > c$ and the ratios $b/a$ and $c/a$ characterize the ellipsoid shape. Specifically, spherical halos have $(b/a, c/a)\approx(1, 1)$, elongated (or prolate) halos have $(b/a, c/a)\approx(0, 0)$ while flattened (or oblate) halos have $(b/a, c/a)\approx(1, 0)$. In Fig. \ref{fig:shape-distribution} we plot the distribution of $(b/a, c/a)$ for halos from both a real HR snapshot and a generated SR snapshot. The distributions were computed using the Gaussian kernel density estimation method from the scipy Python library. We find the SR halo shape distribution is shifted toward lower values of $(b/a, c/a)$, indicating a tendency toward more prolate structures compared to HR halos. To compute the statistical significance of this difference, we performed a permutation test using the $L^2$ distance between the distributions as the test statistic. The observed $L^2$ difference was 0.0044, which was significant at the $\alpha = 0.5\%$ level, with a $p$-value of 0.2\%.

\begin{figure}
        \centering
        \includegraphics[width=1\linewidth]{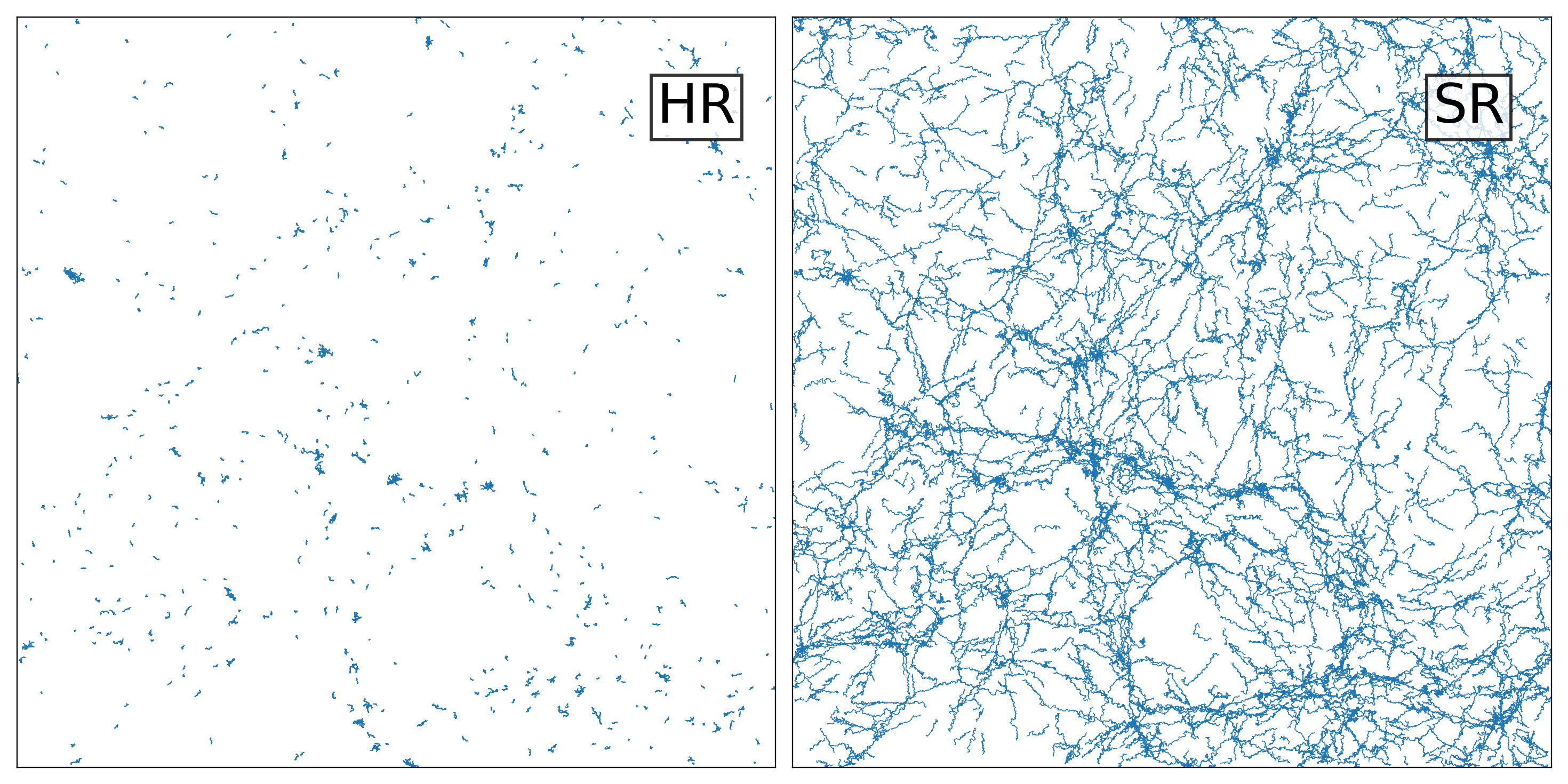}
        \caption{\normalsize Projections of the MSTs for the HR and SR data are shown on the left and right respectively. After pruning and separating, the HR MST consists of 521 connected components, all of which represent nodes of the cosmic web. In contrast, the SR MST has 1017 connected components that exhibit prominent branches that still extend along the filaments. \textit{Takeaway}: The SR MST retains extended branches even after applying a pruning criterion that removes them from the HR MST.}
        \label{fig:pruned-mst}
    \end{figure}

\begin{figure}
        \centering
        \includegraphics[width=1\linewidth]{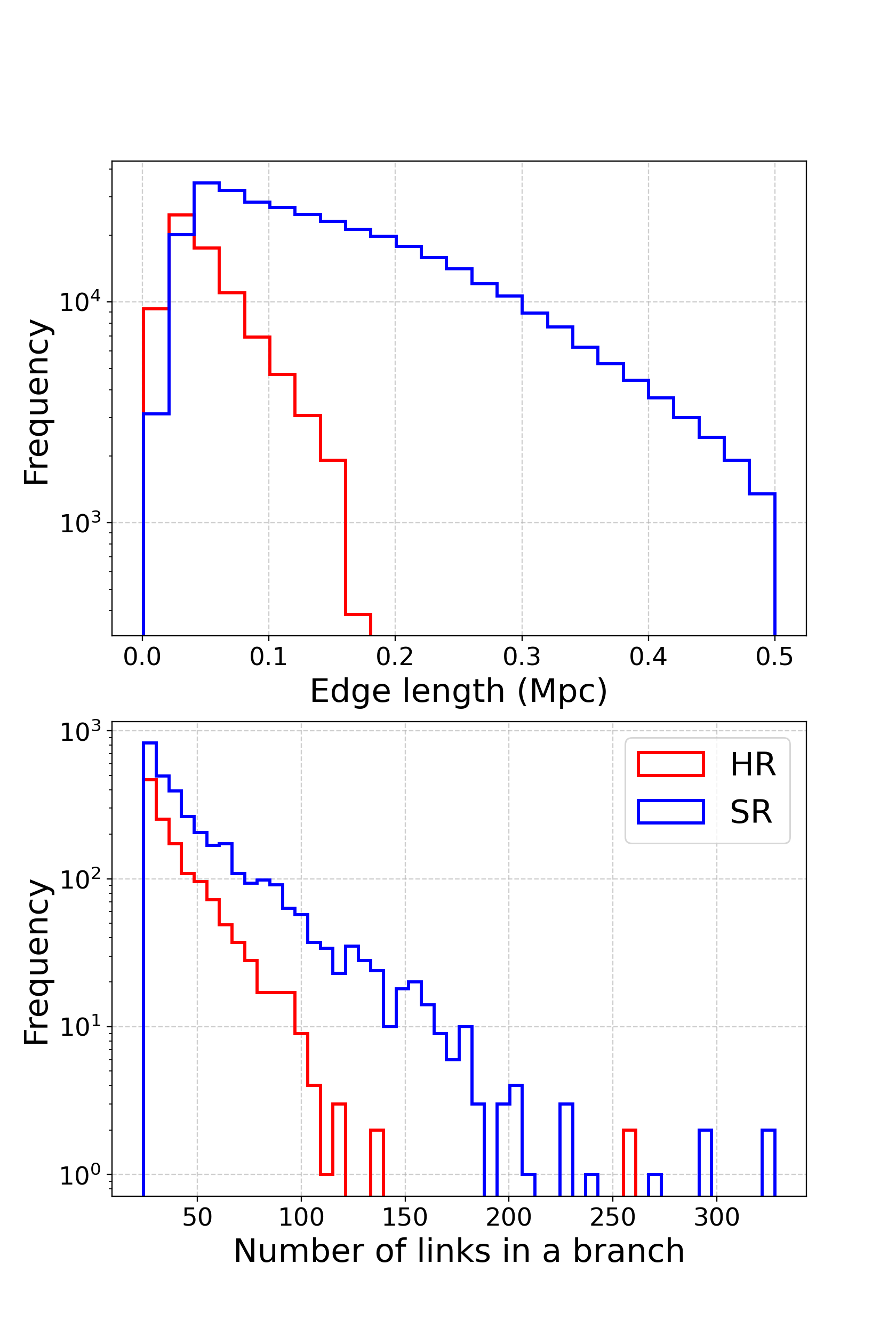}
        \caption{\normalsize The distribution of edge lengths for both HR and SR MSTs are shown in the top panel. The bottom panel shows the distribution of the number of edges in a branch for both MSTs. \textit{Takeaway}: In comparison to the HR MST, the SR MST exhibits longer edge lengths and branches consisting of a greater number of edges.}
        \label{fig:mst-stats}
    \end{figure}

    This bias towards overly smooth dark-matter distributions also affects the shape of the large-scale cosmic web. As seen in Fig. \ref{fig:dark-matter-scatter}, the smearing effect appears to erase much of the filamentary substructure, replacing it with a more diffuse distribution of dark matter along the filaments. To quantify this morphological difference between real and generated data, we analyzed variations of the minimal spanning tree (MST) \citep{Barrow_85, Naidoo_2020} statistic. MSTs were constructed from the dark-matter particle positions in both HR and SR snapshots using the Kruskal algorithm as implemented in scipy. The MSTs were also pruned of all branches with 25 or fewer links. Any edges longer than 3 times the average edge length were also removed. This level of pruning was enough to remove almost all of the branches extending along the filaments of the HR data, leaving dense clusters of MST nodes corresponding to the nodes of the cosmic web. In contrast, large branches corresponding to filaments persist in the SR data after applying the same pruning criteria. Projections of the pruned MSTs for both cases are shown in Fig. \ref{fig:pruned-mst}. We also present the edge length distributions of the MSTs in the top panel of Fig. \ref{fig:mst-stats} and the distribution of the number of edges in a branch in the lower panel. Compared to HR data, SR data exhibits both longer edge lengths and branches composed of more edges. Both of these are indicative of a more diffuse distribution of dark-matter particles along the cosmic web, particularly along the filaments. These results reinforce the conclusion that the SR method struggles to accurately reproduce the geometry of the dark-matter distribution.

    This smearing effect is consistent with known limitations in the SR literature where some GAN-based models can produce outputs that are biased toward overly smooth solutions \citep[e.g.,]{Wang_2023}. While in some cases this bias can be attributed to certain loss functions \citep[e.g.,]{Blau_2018, Wang_2019}, a key issue here is that the output of our WGAN model is inherently under-constrained on scales below the resolution of the training targets (such as the cloud-in-cells density field). To empirically test whether increasing the spatial resolution of the discriminator input could mitigate this effect, we trained two additional models using the generalized critic architecture described in Section \ref{sec:critic-model}. These models used density fields with spatial dimensions of $68^3$ and $140^3$, corresponding to one and two additional residual blocks in the density branch, respectively. The base number of channels for the density branch in both cases was set to $16$ while it was kept at $64$ for the main branch. A comparison of dark-matter distributions generated by these models is shown in Fig \ref{fig:level_comparison} in the appendix. We found that increasing the resolution of the density field in this way did not remove the smearing effect. Moreover, the wall-clock time per training epoch increased by factors of 2 and 140, respectively. Further increases in resolution were not tested, as the required memory exceeded the capacity of the H100 GPU used for this work.

%-----------------------------------------------------------------
\section{Conclusions and Outlook}\label{sec:conclusion}
    \noindent In this work we trained several WGAN models to enhance the particle resolution of snapshots from dark-matter-only simulations, following the method outlined in \cite{Li_2021}. While we found that these models could produce data with summary statistics that agree well with real high-resolution (HR) simulations, we identified a notable limitation: the shapes of dark matter halos in the super-resolved (SR) data are biased by smeared features not present in the real HR data. This discrepancy in halo morphology, visible in both scatter plots and halo shape metrics, suggests that while global statistics may be preserved, important small-scale structural details are lost. This limitation must be considered when using SR models for scientific analyses that are sensitive to halo shapes.
    
    Despite these limitations, SR methods remain a valuable tool for cosmological studies. Given the impressive demonstrations of the techniques used in \cite{Zhang_2024} and \cite{Jacobus_2024}, it is tempting to judge these methods as cheap replacements for HR simulations. Nevertheless, it is important to recognize that these models are approximating the physics occurring on length scales that are not resolved by the LR simulation. Therefore, rather than being viewed as a replacement for HR simulations, SR methods should be regarded as simulation-informed subgrid models. As such, the value of these methods stems from their ability to help bridge the gap between LR and HR simulations when computational resources are limited.
    
    Future work could address the issue of smeared features by developing additional loss terms that account for physical properties beyond summary statistics. For example, quantifying the divergence of halo shapes from those appearing in HR data could provide a regularization term to discourage the WGAN from generating overly smoothed solutions. To this end, the shape parameters used in this work or other metrics, such as Minkowski functionals, that encode information about dark matter halos could be useful. This approach aligns with the methods used by \cite{Helfer_2024} to correct numerical relativity simulations that break conservation laws during mesh refinement. Exploring alternative generative frameworks, such as stable diffusion models, may also provide insights into whether similar limitations persist or if new architectures can overcome this issue. A detailed comparison between WGANs and stable diffusion methods in this context could help identify the strengths and weaknesses of each approach.

    Applications of SR techniques are significant, and we will explore some of those applications in future works. One example would be the ability to extend the resolution of cosmological N-body simulations to smaller scales opens the possibility of better tracking the assembly history of halos at a relatively small cost. In addition, to the positional information used in this work, this would require temporal (redshift) information and kinematics - however both of these extensions are tractable. Gaining insight into the assembly history of halos at high redshift would be particularly useful for understanding the formation of (rare) early cosmic structures which require both large volume statistics and high resolution \citep[e.g.,][]{Trinca_2022}. These smaller halos (atomic cooling halos and molecular cooled halos) have the potential to host the first stellar objects in our Universe. Semi-analytical models, which use merger trees as their backbone, could benefit hugely from SR techniques \citep[e.g.,][]{Spinoso_2023}. 
    
    A second example is the use of SR techniques to capture small-scale structure within simulations of large-scale structure spanning several gigaparsecs. The shortcoming identified with the current SR method may not be problematic for certain applications in this direction. For instance, neutral atomic hydrogen (HI) 21-cm intensity mapping \citep[e.g.,][]{VN_2018} or modeling the growth of the ionizing background during reionization \citep[e.g.,][]{Greig_2022}, since the HI density field is effectively smoothed on scales of $\sim$10 Mpc. The ability to rapidly generate matter density fields - and, by extension, galaxy density fields informed by high-resolution hydrodynamical simulations - will be an invaluable tool for cosmology with the SKA telescope \citep[e.g.,][]{Maartens_2015}.

\section{Code and Data Access}
    \noindent All neural network models and scripts used to produce the results in this paper were implemented in Python and PyTorch \citep{Paszke2019-hd} and are available at \url{https://github.com/brenjohn/DMSR-WGAN} on the "on-wgan-super-resolution" branch. Steps for reproducing results are also available there along with model weights used for this work. For recreating the datasets, the parameter files used for generating the initial conditions and running the dark-matter-only simulations, from which the datasets were created, are also provided. The work presented here made use of the numpy, scipy, networkx and matplotlib Python libraries \citep{harris2020array, 2020SciPy-NMeth, SciPyProceedings_11, Hunter:2007}.

\section*{Acknowledgments}
    JB and JR acknowledge support from the Royal Society and Research Ireland through the University Research Fellow programme under grant number URF$\backslash$R1$\backslash$191132.
    JR acknowledges support from the Research Ireland Laureate programme under grant number IRCLA/2022/1165. CP and SB acknowledge the Australian Research Council (ARC) Centre of Excellent for All Sky Astrophysics in 3 Dimensions (ASTRO 3D; project \#CE170100013). Parts of this research was undertaken with the assistance of resources from the National Computational Infrastructure (NCI Australia), an NCRIS enabled capability supported by the Australian Government. SB acknowledges support from grant PID2022-138855NB-C32 funded by MICIU/AEI/10.13039/501100011033 and  ERDF/EU.
    YQ is supported by the ARC Discovery Early Career Researcher Award (DECRA) through fellowship \#DE240101129.
    We acknowledge EuroHPC Joint Undertaking for awarding us access to MareNostrum5 hosted by BSC in Spain. The research in this paper made use of the SWIFT open-source simulation code (http://www.swiftsim.com, Schaller et al. 2018) version 1.0.0.

\bibliographystyle{mn2e}
\bibliography{oja_template}

\newpage
\begin{appendix}

    \section{Appendix A: Results from Generalized Critic Models}
    \label{ap:level_comparison}
\begin{figure}[h]
        \centering
        \includegraphics[width=1\linewidth]{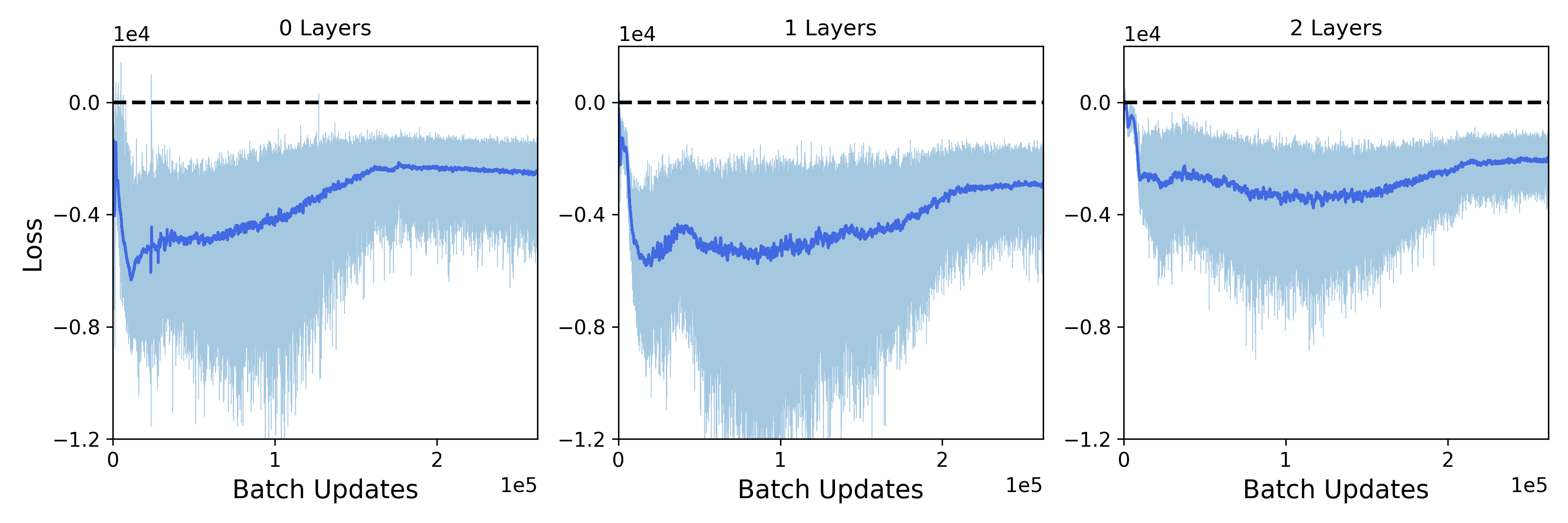}
        \caption{\normalsize Here we show the loss curves for the three WGANs with 0, 1 and 2 additional residual layers in their critic's density branch. The loss curve is shown in light blue while a moving average with a window size of 500 is shown in dark blue.}
        \label{fig:loss_curves}
    \end{figure}

\begin{figure*}[h]
        \centering
        \includegraphics[width=1\linewidth]{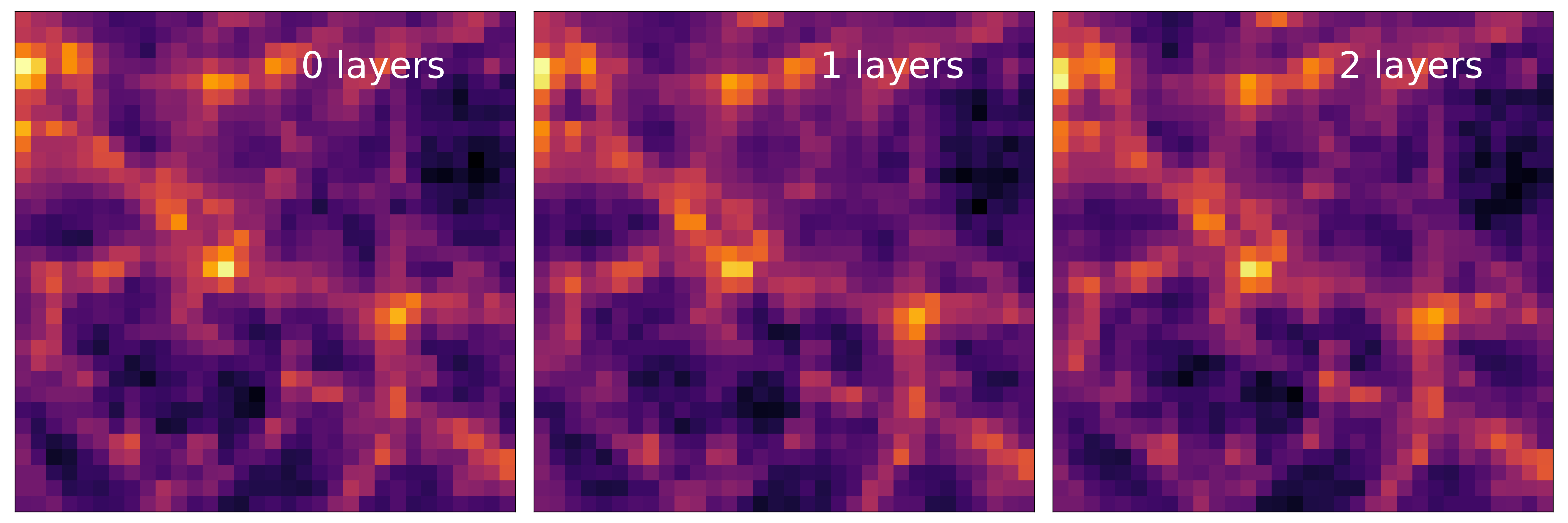}
        \includegraphics[width=1\linewidth]{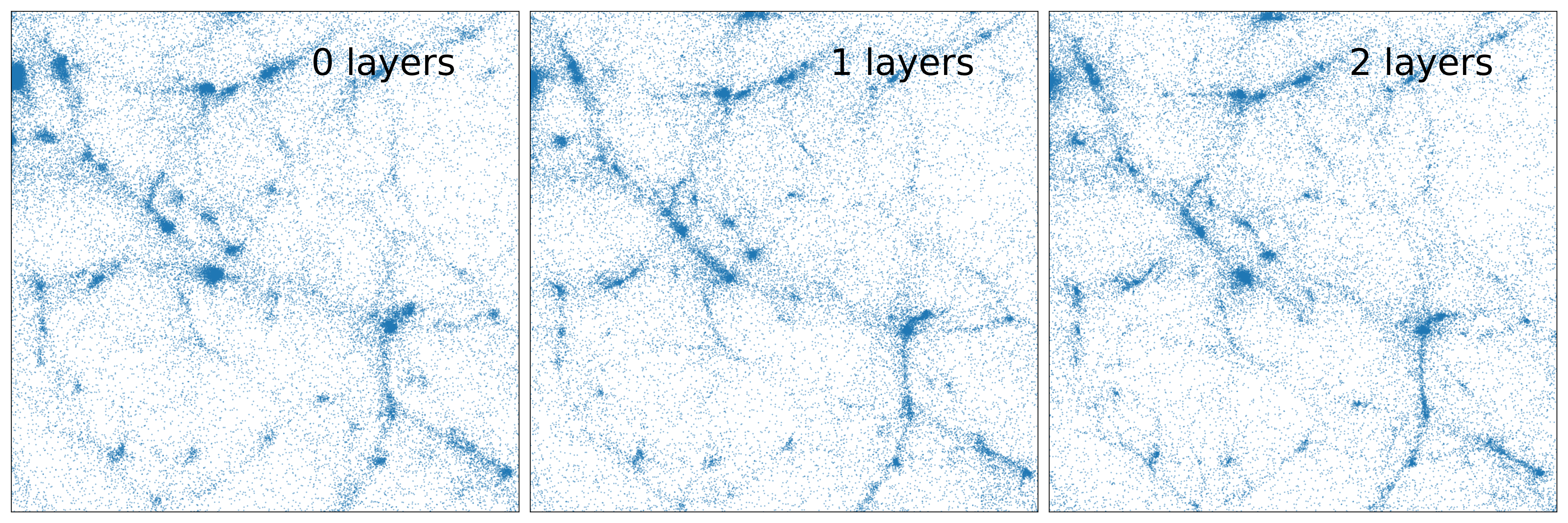}
        \caption{\normalsize Here we show a comparison of the same 25 $\rm{h^{-1}}$ Mpc region shown in Fig. \ref{fig:dark-matter-density}, enhanced by three different WGAN models using different resolution levels to train the critic model. From left to right, the models used had 0, 1 and 2 additional layers in the density branch of the critic model as described in the methodology. As can be seen, the presence of smeared features persists even when the resolution of the density field has been increased by 2 levels. Although the smeared features become sharper, indicating that the effect is being reduced, training WGAN models with higher density resolution quickly becomes intractable. \textit{Takeaway}: Simply increasing the resolution of the density field used to train the WGAN is not a viable approach to address smearing in enhanced snapshots.}
        \label{fig:level_comparison}
    \end{figure*}

    \noindent Here we include some additional results from WGANs with the generalized critic architecture described in section \ref{sec:critic-model}. Three WGANs were trained whose critic models had 0, 1 and 2 additional residual layers in their density branches. The WGANs were therefore trained using density fields of 25 $\rm{h^{-1}}$ Mpc cubic patches with $32^3$, $68^3$ and $140^3$ cells respectively. Loss curves for the WGANs are shown in Fig. \ref{fig:loss_curves} where the loss function is given by equation \ref{eqn:loss_function}. A comparison of outputs from the three models is shown in Fig. \ref{fig:level_comparison}. Note that in Wasserstein GANs, the absolute value of the critic loss approximates the Wasserstein distance between the generator's learned distribution and the true distribution of the training data. Therefore, lower (i.e., closer to zero) critic loss values generally indicate a smaller discrepancy between the two distributions. For a detailed discussion, we refer the reader to \citet{Gulrajani_2017}. However, it is important to emphasize that the critic loss is not typically used as a direct measure of generator quality. Consequently, loss values are not used for model selection or early stopping in WGAN training. Instead, critic and generator loss curves serve as diagnostic tools to monitor training dynamics and assess stability. We include these loss curves in the appendix to assist readers who may wish to reproduce our results or compare training behavior.

\end{appendix}

\end{document}